\documentclass[twocolumn,showpacs,preprintnumbers,amsmath,amssymb]{revtex4}

\usepackage[russian]{babel}

\usepackage{graphicx}
\usepackage{graphics}
\usepackage{dcolumn}
\usepackage{bm}

\usepackage{latexsym}
\usepackage{amssymb}
\usepackage{amsmath}

\begin{document}                  
\title{Dzyaloshinskii--Moriya interaction: How to measure its sign in weak ferromagnetics?}

\author{Vladimir E.~Dmitrienko\/\footnote{dmitrien@crys.ras.ru},
Elena~N.~Ovchinnikova$^{\dagger}$, Jun~Kokubun$^{\ddagger}$,
Kohtaro~Ishida$^{\ddagger}$}

\address{A.V.~Shubnikov Institute of Crystallography, 119333
Moscow, Russia\\
$^{\dagger}$Department of Physics, Moscow State University,
Moscow, Russia\\
$^{\ddagger}$Faculty of Science and Technology, Tokyo University
of Science, Noda, Chiba 278-8510, Japan}


\begin{abstract}
Three experimental techniques sensitive to the sign of the
Dzyaloshinskii--Moriya interaction are discussed: neutron
diffraction, M\"ossbauer $\gamma$-ray diffraction, and resonant
x-ray scattering. Classical examples of hematite
($\alpha$-Fe$_2$O$_3$) and MnCO$_3$ crystals are considered in
detail.

\end{abstract}


\maketitle

Weak ferromagnetism (WF) of antiferromagnetics is a classical
example of an initially small and controversial physical problem
that later produces a strong impact on the general picture of
magnetic phenomena. From the very beginning, the modern
theoretical consideration of WF developed by Dzyaloshinskii and
Moriya was based on symmetry arguments, both phenomenological
\cite{Dzyaloshinskii57,Dzyaloshinskii58} and microscopic
\cite{Moriya60a,Moriya60b}. It was shown that appropriate crystal
symmetry allows the following term in the interaction of two
antiferromagnetic sublattices $\bm S_1$ and $\bm S_2$
\begin{equation}
\bm D \cdot [\bm S_1\times\bm S_2], \label{Dvector}
\end{equation}
which favors to (usually small) canting angle between $\bm S_1$
and $\bm S_2$; here $\bm D$ is a vector parameter of the
Dzyaloshinskii--Moriya interaction. Possible directions of $\bm D$
were found \cite{Moriya60b} for different local symmetries.
Significant progress was recently achieved in \emph{ab initio}
calculations of $\bm D$ (see \cite{Mazurenko05} and references
therein). Besides fundamental interest, the Dzyaloshinskii--Moriya
interaction is a very important ingredient of magnetoelectric
effects with possible applications to spintronics.

The canted spin arrangement is just responsible for WF. Both the
magnitude of WF and the canting angle are proportional to $|\bm
D|$ and therefore it seems that the sign of $\bm D$ is not
important at all. According to Eq. (\ref{Dvector}), the sign
obviously depends on our choice which of sublattices is 1 or 2 and
therefore it is usually claimed that the sign is conventional.
From the phenomenological point of view this is true because in
macroscopic theory the phase of antiferromagnetic arrangement is
not fixed relative to the crystal lattice. However at the atomic
level the phase can be fixed owing to the Dzyaloshinskii--Moriya
interaction and the sign of this interaction is crucial for
relation between the local crystal structure and magnetic
ordering. For instance, this sign determines the handedness of
spin helix in crystals with the noncentrosymmetrical B20 structure
\cite{Nakanishi80,Bak80}. In this paper we show how one can
measure it in classical WF crystals like $\alpha$-Fe$_2$O$_3$ or
MnCO$_3$.

Let us rewrite Eq. (\ref{Dvector}) in a more invariant form not
depending on any arbitrary choice of sublattices. If two atoms
with spins $\bm s_1$ and $\bm s_2$ are located at the points $\bm
r_1$ and $\bm r_2$, then we can add the following scalar to the
energy of their interaction
\begin{equation} T_{jkm}s_{1j}s_{2k}(\bm r_1-\bm r_2)_m,
\label{Ttensor}
\end{equation}
where an antisymmetric tensor, $T_{jkm}=-T_{kjm}$, characterizes
interaction of spins $\bm s_1$ and $\bm s_2$ through intermediate
crystal matter. The properties (in particular symmetry) of the
intermediate matter determines the properties of this tensor
including symmetry restrictions on its tensor components. It is
well known also that any third-rank antisymmetric tensor is
equivalent to a second-rank pseudo-tensor $A_{nm}$:
$T_{jkm}=\epsilon_{jkn}A_{nm}$ where $\epsilon_{jkn}$ a unitary
antisymmetric pseudo-tensor ($A_{nm}$ changes its sign under
inversion). The relation between $\bm D$, $A_{nm}$ and $T_{jkm}$
is given by
\begin{equation} D_n=\frac12\epsilon_{jkn}T_{jkm}(\bm r_1-\bm
r_2)_m=A_{nm}(\bm r_1-\bm r_2)_m. \label{TtoD}
\end{equation}

Using the well known symmetry restrictions on the third-rank
antisymmetric tensors \cite{Sirotin75} we can obtain from Eq.
(\ref{Ttensor}) all the symmetry restrictions on $\bm D$ found in
\cite{Moriya60b}. In particular, $A_{nm}=0$, $T_{jkm}=0$ and
$D_{m}=0$ if the points $\bm r_1$ and $\bm r_2$ are related by
inversion symmetry (rule 1 from \cite{Moriya60b}). If there is an
$n$-fold rotation axis ($n\ge 2$) along $\bm r_1-\bm r_2$ then
$\bm D$ is parallel to $\bm r_1-\bm r_2$ (rule 5 from
\cite{Moriya60b}), \emph{etc.}

However, there is an important principal difference between $\bm
D$ and $T_{jkm}$: tensor $T_{jkm}$ (or pseudo-tensor $A_{nm}$) can
be considered as a field on the lattice, it should be invariant
relative to all the symmetry operations of the space group. In
particular, it is determined by the same parameters at all
equivalent lattice points; of course, one should take into account
corresponding crystallographic operations connecting those
equivalent points: rotations (changing orientations of the
principal axes) and space inversions (changing signs of all
components of $T_{jkm}$). On the contrary, the pattern of vector
$\bm D$ on the lattice cannot be obtained by pure crystallographic
operations and some additional consideration is needed (see
discussion of La$_2$CuO$_4$ in \cite{Coffey90,Coffey91}).

We conclude this short introduction with a remark that Eq.
(\ref{Ttensor}) cannot be used for quantitative description of WF;
modern first-principles theoretical considerations are more
appropriate \cite{Mazurenko05}. Nevertheless this expression can
be used for better understanding of symmetry aspects of the
problem and now we will show that this is really the case.

The appearance of the antisymmetric third-rank tensor suggests an
idea that there is some chiral effect behind this. In fact, it is
known for a long time \cite{Dzyaloshinskii64} that the
Dzyaloshinskii--Moriya interaction can produce long-period
magnetic spiral structures in ferromagnetic and antiferromagnetic
crystals lacking inversion symmetry. This effect was suggested for
MnSi and other crystals with B20 structure
\cite{Nakanishi80,Bak80} and it has been carefully proved that the
sign of the Dzyaloshinskii--Moriya interaction, hence the sign of
the spin helix, is determined by the crystal handedness
\cite{Grigoriev09}. More delicate situation with chirality occurs
in typical WF crystals with $R\bar{3}c$ symmetry
($\alpha$-Fe$_2$O$_3$, MnCO$_3$, \emph{etc.}) which are
centrosymmetric.

At first let us consider classical WFs, carbonates of transition
metals, for instance MnCO$_3$ \cite{Landau82}. In its primitive
rhombohedral unit cell with the space group $R\bar{3}c$, there are
two Mn atoms at crystallographically equivalent inversion centers
$\bar{3}$, vertices ($0,0,0$) and body-centers
($\frac12,\frac12,\frac12$). Atoms at the vertices and
body-centers have almost opposite magnetic moments lying in the
planes normal to the threefold axis. According to the crystal
symmetry the moments should not be exactly opposite and in fact
the moments are slightly canted so that the resulting WF moment is
along one of three twofold axes. There are also two carbon atoms
at points ($\frac14,\frac14,\frac14$) and
($\frac34,\frac34,\frac34$).

The physical origin of WF in  MnCO$_3$ is a weak relativistic
interaction between spins in the lattice with the $R\bar{3}c$
space group. But what is the structural origin of this $R\bar{3}c$
symmetry? If we consider only Mn atoms, the symmetry of lattice
would be $R\bar{3}m$ and WF would be impossible. This symmetry
will not change if the carbon atoms are taken into account. Only
oxygen atoms change $R\bar{3}m$ symmetry to $R\bar{3}c$; thus
their configuration is crucial for the value (and sign) of the
Dzyaloshinskii--Moriya interaction and it is worthy of a more
careful consideration.

In  MnCO$_3$, there are six hexagonal Mn layers per the lattice
period along the threefold axis, so that atoms of the next layer
is just under the centers of triangles formed by atoms from the
previous layer, and layer sequence is $ABCABC\ldots$, like in the
\emph{fcc} lattice. These equidistant layers has $z$-coordinates
equal to $z=0,\frac16,\frac13,\frac12,\frac23,\frac56$; in each
layer all the spins are parallel and are lying in the layer plane.
The spin of neighboring layers are almost opposite. Here and below
the standard hexagonal setting of the rhombohedral lattice is used
\cite{TablesA}.

Considering the first two layers one can see that between them, at
$z=\frac{1}{12}$, there is a layer of oxygen and carbon atoms; the
point symmetry of this layer is 32, just because of low symmetry
of oxygen positions: oxygen atoms are at the $18e$ positions with
point symmetry 2 and with coordinates equivalent to
$(x_O,0,\frac14)$ where $x_O\approx 0.27$. Carbon atoms are at the
$6a$ positions $(0,0,\frac14)$ with point symmetry 32.

It is very important that this intermediate layer is
noncentrosymmetric and therefore both vector $\bm D$ and tensor
$T_{jkm}$ can have some nonzero values. For pairwise interaction
of spins from different layers, tensor $T_{jkm}$  has symmetry 1,
but being averaged over all pairs it has of course the symmetry of
the intermediate layer. For symmetry 32, tensor $T_{jkm}$ is
determined by two independent parameters, say $T_{yzx}$ and
$T_{xyz}$ \cite{Sirotin75}, but only the latter leads, according
to Eq. (\ref{Ttensor}), to a twist angle between the Mn spins
lying in the first and second layers. The sign of this twist angle
is just the sign of $T_{xyz}$. This twist violates the right-left
symmetry and, in a figurative sense, we can say that the
intermediate C-O layer is chiral.

\begin{figure}
\includegraphics[width=\columnwidth]{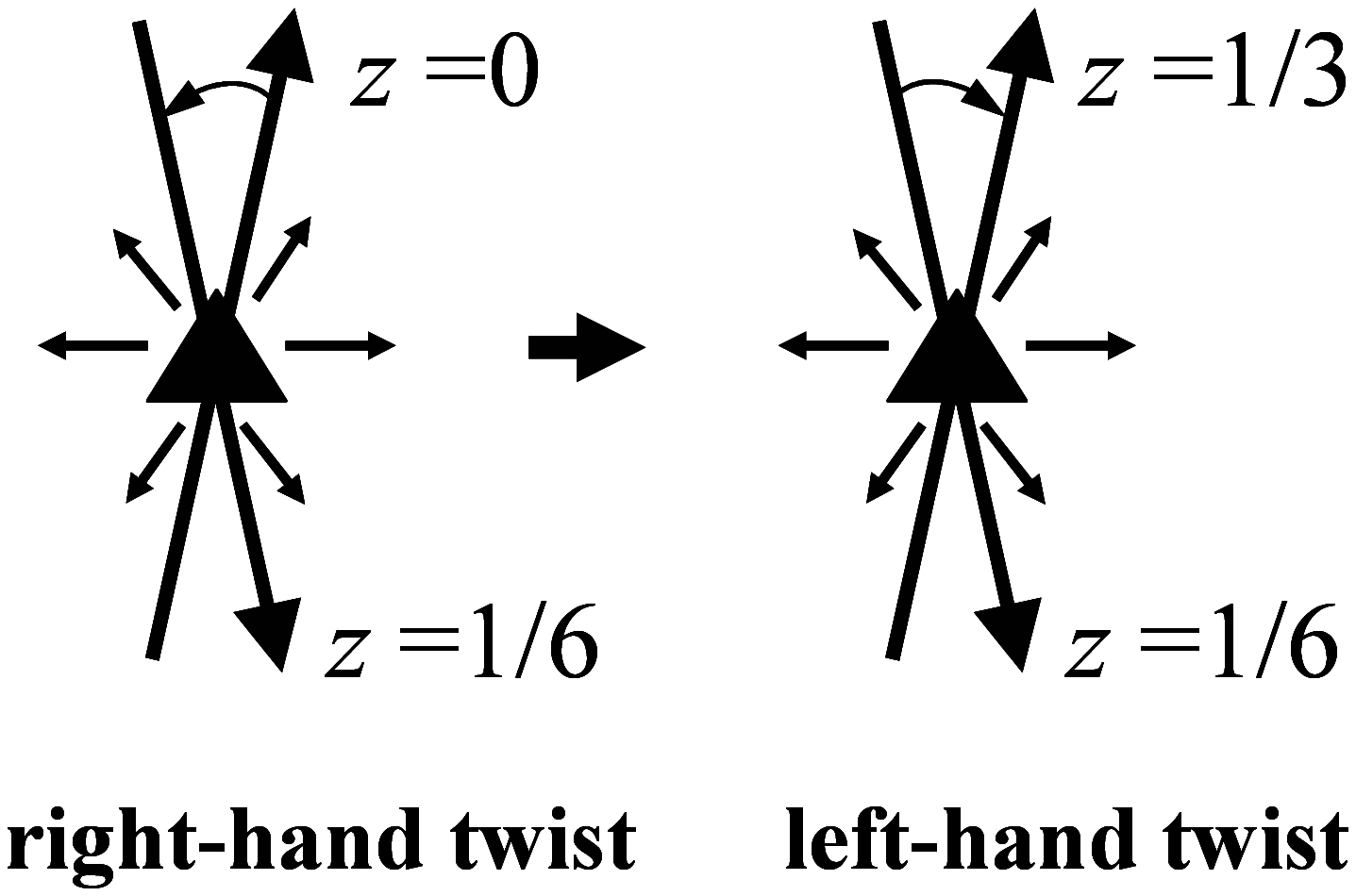}
\caption{Right-hand and left-hand twists of moments between layers
alternating along $z$ axis. Triangles indicate the threefold axis
normal to the figure plane; all possible directions of WF moments
in MnCO$_3$ (twofold axes) are shown by small arrows. Big arrows
are spin directions in neighboring layers at different $z$ levels
for the case when external magnetic field is applied in horizontal
direction and DM is positive. A bold small arrow indicates the
direction of WF moment.} \label{figcant}
\end{figure}

The next intermediate C-O layer is at $z=\frac{1}{4}$, i.e.
between the Mn layers at $z=\frac{1}{6}$ and $z=\frac{1}{3}$, and
for this layer all the components of tensor $T_{jkm}$ change sign
due to inversion centers at $z=\frac{1}{6}$. Thus the
"chirality"\, of this layer is opposite to that of the first
intermediate layer, hence the small twist angle between spins is
also opposite and for the Mn layer at $z=\frac{1}{3}$ the spin
orientation exactly coincides with the spin orientation for the Mn
layer at $z=0$ (see Fig.~\ref{figcant}). And then this repeats
from layer to layer. We see that two alternating local twists,
left and right, between neighboring Mn layers result in
macroscopic canting angle between magnetic sublattices in
centrosymmetric WF crystals.

If we change the sign of $x_O$ so that $x'_O=1-x_O\approx 0.73$,
then the layer "chirality"\, changes to opposite, $T_{xyz}$ change
the sign and the twist angle also change the sign. From pure
crystallographic point of view both values, $x_O$ and $x'_O$, are
equivalent, they simply correspond to the lattice origin shifted
at a half-period, from (0,0,0) to $(0,0,\frac12)$. In the
crystallographic databases, both values of $x_O$ are cited for
$\alpha$-Fe$_2$O$_3$; nevertheless sometimes the first-principle
results for WF in $\alpha$-Fe$_2$O$_3$ do not indicate which value
of $x_O$ they really adopt (see for instance \cite{Sandratski96}).
We see also that in $R\bar{3}c$ crystals the sign of the
Dzyaloshinskii--Moriya interaction changes to opposite at one
half-period, therefore the idea to measure this sign by the
M\"ossbauer absorption \cite{ozhogin68} cannot be correct.

According to $R\bar{3}c$ symmetry there are six possible
orientations of the WF moments (along plus and minus directions of
three twofold axes; in Fig.~\ref{figcant}, they are shown by short
arrows, like in \cite{TablesA}). Application of an external
magnetic field along one of these directions makes the
corresponding \emph{ferromagnetic} domain energetically favorable.
And if the orientation of the ferromagnetic domain is fixed then
the Dzyaloshinskii--Moriya interaction fixes the phase of
\emph{antiferromagnetic} sequence of moments in this domain
\cite{Nathans64}.

Now we are ready to consider the main item of this paper: how to
measure the sign of the Dzyaloshinskii--Moriya interaction in WFs?
First of all, a strong enough magnetic field should be applied to
obtain the single domain state where the Dzyaloshinskii--Moriya
interaction pins antiferromagnetic ordering to the crystal
lattice. Next, \emph{single-crystal} diffraction methods sensitive
both to oxygen coordinates and to the phase of antiferromagnetic
ordering should be used. In other words, one should observe those
Bragg reflections $hk\ell$ where \emph{interference} between
magnetic scattering on Mn atoms and nonmagnetic scattering on
oxygen atoms is significant. There are three suitable techniques:
neutron diffraction, M\"ossbauer $\gamma$-ray diffraction, and
resonant x-ray scattering. We will discuss now their advantages
and disadvantages.

Because of the layered magnetic structure alternating along
$z$-axis, the reflections with strong magnetic scattering
correspond to the reciprocal lattice vectors $\bm H_{hk\ell}$ with
odd $\ell$. At first let us consider scattering on oxygen atoms;
the expressions for the oxygen structure amplitudes $F_{ox}(\bm
H)$ are similar for all three techniques and looks like
($\ell=2n+1$):
\begin{eqnarray}
F_{ox}(\bm H)=2A_{ox}[\cos2\pi(hx_{ox}+\ell/4)+\cos2\pi(kx_{ox}+\ell/4) \nonumber \\
+\cos2\pi(hx_{ox}+kx_{ox}-\ell/4) \nonumber  \\
=8A_{ox}(-1)^{n+1}\sin\pi hx_{ox}\sin\pi
kx_{ox}\sin\pi(h+k)x_{ox}, \label{FO}
\end{eqnarray}
where $A_{ox}$ is proportional to the oxygen atomic scattering
factor (for x-rays) or to the nuclear scattering length (for
neutrons); it is practically real because oxygen absorption is
very small for thermal neutrons or hard x-rays. There is no
contribution from carbon \cite{TablesA}. Here and below some
factors (such as atomic and magnetic formfactors, the
Debye--Waller factor, \emph{etc.}) are omitted because
corresponding expressions are well known and implemented into
routine computer programs used for diffraction experiments. It is
clear from this equation that one should measure reflections with
$hk(h+k)\ne 0$.

In the case of neutron diffraction, one can adopt the standard
technique using the polarization ratio $R$, i.e. the ratio of
reflection intensities for incoming neutrons with spin $\bm\sigma$
parallel and antiparallel to the direction of applied magnetic
field. For $\ell=2n+1$ this ratio is given by the following
expression containing interference between nuclear scattering by
oxygen atoms and magnetic scattering by Mn atoms
\begin{eqnarray} R(hk\ell)=\frac{|F_{ox}(\bm H)+
\bm\sigma\cdot\bm Q(\bm H)|^2}{|F_{ox}(\bm H)-\bm\sigma\cdot\bm
Q(\bm H)|^2} \label{R1}\\
=\frac{|F_{ox}(\bm H)- s_{DM}A_{mag}k(2h+k)|^2}{|F_{ox}(\bm
H)+s_{DM}A_{mag}k(2h+k)|^2}, \label{R2}
\end{eqnarray}
where $\bm Q(\bm H)$ is the magnetic structure amplitude for
reflection $\bm H$, $\bm Q(\bm H)\propto \bm M_{\bm H}-\bm H(\bm
H\cdot\bm M_{\bm H})/\bm H^2$, $\bm M_{\bm H}$ is the
correspondent Fourier harmonic of the vector field describing the
electron-magnetization distribution, $s_{DM}$ is the sign of the
Dzyaloshinskii--Moriya interaction between the first two layers of
Mn atoms and $A_{mag}$ includes all routine factors describing
neutron magnetic scattering. Calculating (\ref{R2}) from
(\ref{R1}) we took into account the geometry shown in
Fig.~\ref{figcant} (i.e. $\bm \sigma$ is directed horizontally,
$\bm M_{\bm H}$ is directed vertically so that $\bm\sigma\cdot\bm
M_{\bm H}=0$ if small canting is neglected, \emph{etc.}). In
particular, factor $s_{DM}$ appears just because the phase of
antiferromagnetic sequence is fixed by the Dzyaloshinskii--Moriya
interaction. All other factors in Eq. (\ref{R2}) are more or less
known and one can determine $s_{DM}$ from rather rough
measurements giving $R(hk\ell)<1$ or $R(hk\ell)>1$. Notice that
there is an additional condition for this measurements, $2h+k\ne
0$, which appears in Eq. (\ref{R2}) from $\bm H\cdot\bm M_{\bm
H}\ne 0$. To avoid confusion we should notice that expression
(\ref{FO}) obeys threefold symmetry whereas magnetic scattering
does not, because of the external field applied perpendicular to
the threefold axis.

The technique of polarized neutron diffraction was used for
measurements of the sign of small angular deviations of moments in
MnF$_2$ \cite{Brown81} but in that case the deviation is
introduced by the single-spin anisotropy \cite{Moriya60}. To the
best of our knowledge, there were no attempts to measure with this
technique the sign of the Dzyaloshinskii--Moriya interaction.

In hematite ($\alpha$-Fe$_2$O$_3$), the situation is slightly more
complicated because iron atoms are in the positions ($0,0,z_{Fe}$)
with the point symmetry 3 and neighboring Fe layers are coupled
either ferromagnetically or antiferromagnetically. Like in
MnCO$_3$, the $R\bar{3}c$ symmetry is induced by oxygen atoms. The
Fe layers coupled feromagnetically are related by inversion and,
according to the Morya rules, there is no canting between them
($T_{jkm}=0$ for inversion centers). The antiferomagnetic
neighboring layers interact via an oxygen layer with symmetry 32
and alternating right and left twists of their moments lead to a
macroscopic WF moment. Eq. (\ref{R2}) transforms to
\begin{eqnarray} R(hk\ell)=
\frac{|F_{ox}-s_{DM}A_{mag}k(2h+k)\cos2\pi\ell z_{Fe}|^2}
{|F_{ox}+s_{DM}A_{mag}k(2h+k)\cos2\pi\ell z_{Fe}|^2}. \label{R3}
\end{eqnarray}
The additional factor $\cos2\pi\ell z_{Fe}$ (where $z_{Fe}\approx
0.355$) allows us to change the value and sign of magnetic
scattering just changing $\ell$. A suitable reflection (210
rhombohedral, i.e. $2\bar{1}3$ hexagonal) had been studied in
\cite{Nathans64}; however, as it was noted in that paper, the
result was controversial: in the great majority of observations,
$R(2\bar{1}3)=1/R(\bar{2}1\bar{3})$ whereas
$R(2\bar{1}3)=R(\bar{2}1\bar{3})$ would be expected from the
symmetry of hematite. Thus we cant extract the sought sign of the
Dzyaloshinskii--Moriya interaction and more careful experiments
are needed.

It should be again emphasized that our symmetry-based arguments
are only qualitative: the \emph{ab initio} calculations for
$\alpha$-Fe$_2$O$_3$ show that the torque induced by neighboring
antiferomagnetic layer is opposite to the total torque
\cite{Mazurenko05}. This confirms importance of suggested direct
measurements of the sign of the total torque.

The M\"ossbauer diffraction can be used in a similar way. In this
case, there is no need to vary the photon polarization and one can
study intensity $I_{hk\ell}(E)$ of reflections as a function of
photon energy $E$:
\begin{eqnarray}
I_{hk\ell}(E)=|F_{ox}(\bm H)- s_{DM}B_{mag}(\bm H,E)|^2,
\label{Mossbauer}
\end{eqnarray}
where $B_{mag}(\bm H,E)$ is the magnetic M\"ossbauer structure
factor of the $hk\ell$ reflections with $\ell=2n+1, hk(h+k)\ne 0$.
The function $B_{mag}(\bm H,E)$ is well known \cite{Belyakov75}
and its real part, which interferes with the first term in
(\ref{Mossbauer}), changes sign when $E$ passes through resonances
provided by the hyperfine splitting of nuclear levels. This should
facilitate the observation of interference between the two terms.
For $^{57}$Fe both terms in (\ref{Mossbauer}) may be of the same
order of magnitude. The M\"ossbauer diffraction was observed in
many crystals including WF $\alpha$-Fe$_2$O$_3$ \cite{Smirnov69}
and FeBO$_3$ \cite{Kovalenko77} (an analog of MnCO$_3$) but these
studies were concentrated mainly on pure magnetic scattering
rather then on its interference with scattering on oxygen atoms.
Contrary to neutrons, the M\"ossbauer diffraction can be used for
very thin layers but the number of possible crystals is rather
limited by the list of suitable M\"ossbauer isotopes.

Another promising approaches to the sign measurements can be
related with resonant x-ray diffraction, i.e. diffraction near
x-ray absorption edges. It is sensitive both to structural and
magnetic ordering especially near $L$ absorption edges (see recent
surveys \cite{Lovesey05,Dmitrienko05,Laan08}). However, in $d$
magnetic metals, the $K$ edge is the only appropriate for
diffraction, and for this edge magnetic scattering is several
orders of magnitude smaller than conventional charge scattering by
electrons. Thus, for reflections of $\ell=2n+1, hk(h+k)\ne 0$
type, the intensity is given by eq. (\ref{Mossbauer}) with the
second term much smaller then the first one. Therefore the
reliable observation of interference between two terms will be
perhaps very difficult.

However, resonant x-ray scattering provides another nontrivial way
to measure $s_{DM}$. The asymmetric oxygen environment of
transition metals induces some additional anisotropy of their
non-magnetic scattering amplitude, so that just owing to this
anisotropy the reflections with $\ell=2n+1$ can be excited even if
$hk(h+k)=0$. These reflections do not exist out of the resonant
region and they are referred to as "forbidden reflections". There
is no direct contribution to forbidden reflections from oxygen
atoms, equation (\ref{FO}) gives zero, but the sign of the induced
anisotropy depends on asymmetrical arrangement of the oxygen atoms
and, correspondingly, the sign of the non-magnetic structure
amplitude of forbidden reflections with $\ell=2n+1, hk(h+k)=0$ is
proportional to the sign of $x_O$. For crystals with $R\bar{3}c$
symmetry these reflections were first observed in
$\alpha$-Fe$_2$O$_3$ \cite{Finkelstein92}. Then it was predicted
that there should be some "chiral"\, dipole-quadrupole
contribution to these reflections \cite{Dmitrienko01} and
interference between different contributions (including magnetic
scattering) have been studied in detail for $\alpha$-Fe$_2$O$_3$
and Cr$_2$O$_3$ crystals \cite{Kokubun08}. It was shown that the
azimuthal dependence of reflection intensity could be strongly
influenced by this interference (especially for the weak 009
reflection, see Figs. 11 and 12 from \cite{Kokubun08}) and
orientation of antiferromagnetic moment was determined in
$\alpha$-Fe$_2$O$_3$ from the observed azimuthal dependence
(without external magnetic field). Exactly the same measurements
in orienting magnetic field would allow us to determine the sign
of the Dzyaloshinskii--Moriya interaction.

The only problem with the last method is that we should rely on
the sign of the x-ray anisotropy of iron atoms calculated with
rather sophisticated computer codes. However it was proved
experimentally for Ge \cite{Mukhamedzhanov07} that those codes (we
used FDMNES \cite{Joly01}) are rather reliable. It is worth noting
that resonant x-ray diffraction and M\"ossbauer diffraction are
element sensitive and moreover the former can distinguish orbital
and spin contributions to magnetic moments.

In conclusion, we see that the experiments similar to those needed
for the sign measurements had been already performed (some of them
long time ago) for all three considered techniques. Therefore we
believe that this paper will stimulate these measurements in
different types of the weak ferromagnetics.

This work is partly supported by Presidium of Russian Academy of
Sciences (program 27/21) and by the Russian Foundation for Basic
Research (project 10-02-00768). Discussions with V.I.~Anisimov and
E.I.~Kats are gratefully acknowledged.

\end{document}